# Additional interfacial force in lattice Boltzmann models for incompressible multiphase flows

Q. Li,[1] K. H. Luo,[1] and Y. J. Gao[2]

[1]School of Engineering Sciences, University of Southampton, SO17 1BJ Southampton, United Kingdom

[2]Department of Zoology, University of Cambridge, CB2 3EJ Cambridge, United Kingdom

The existing lattice Boltzmann models for incompressible multiphase flows are mostly constructed with two distribution functions, one is the order parameter distribution function, which is used to track the interface between different phases, and the other is the pressure distribution function for solving the velocity field. In this brief report, it is shown that in these models the recovered momentum equation is inconsistent with the target one: an additional interfacial force is included in the recovered momentum equation. The effects of the additional force are investigated by numerical simulations of droplet splashing on a thin liquid film and falling droplet under gravity. In the former test, it is found that the formation and evolution of secondary droplets are greatly affected, while in the latter the additional force is found to increase the falling velocity and limit the stretch of the droplet.

In the past two decades, the lattice Boltzmann (LB) method for simulating multiphase flows has attracted much attention. Historically, the first multiphase LB model was proposed by Gunstensen *et al.* [1]. Since then, many multiphase LB models have been developed. These models can be generally classified into four categories, i.e., the color-gradient model [1, 2], the pseudo-potential model [3, 4], the kinetic-theory-based model [5, 6], and the free-energy-based model [7, 8]. Some brief reviews of the existing multiphase LB models can be found in Refs. [6, 9]. In this paper, we focus our discussion on the LB models for incompressible multiphase flows. To the best of our knowledge, the first incompressible multiphase LB model is proposed by He *et al.* [10], which is a modified model of their kinetic-theory-based multiphase LB model with some approximations. Usually, in He *et al.*'s model the density ratio cannot be very large due to the numerical instability. In order to enhance the numerical stability at high density ratio, Lee and Lin [11] devised a three-stage stable discretization multiphase LB scheme based on He *et al.*'s work. Inamuro *et al.* [12] have also developed a LB model for incompressible multiphase flows with large density ratio. But in their model a pressure correction process is needed to enforce the divergence-free condition after every collision-stream step.

In the above-mentioned incompressible multiphase models, the interface between different phases is tracked by an order parameter distribution function. However, Zheng *et al.* [13] found that the interface capturing equation (the Cahn–Hilliard equation) was not correctly recovered in these models. To solve this problem, they presented a multiphase LB model that can exactly recover the Cahn–Hilliard equation. Recently, Fakhari and Rahimian [9] pointed out that Zheng *et al.*'s model is restricted to density-matched binary fluids and is unable to simulate multiphase flows with noticeable density differences. To this end, they proposed an incompressible multiphase LB model based on the phase-field theory. In their model, the pressure evolution equation proposed by He *et al.*/Lee and Lin is adopted for solving the velocity field. In this sense, Fakhari and Rahimian's model [9] shares a similar feature with He *et al.*'s and Lee and Lin's models: all of these models are constructed with two distribution functions, the order parameter distribution function and the pressure distribution function, although the evolution equations for the order parameter distribution function are different in these models. In this work, we aim to show that the recovered momentum equation of these models is inconsistent with the target one: an additional force is included in the recovered momentum equation. Firstly, we briefly introduce the phase-filed theory.

In the phase-field theory, the thermodynamic behavior of a fluid is expressed by a free energy that is functional of the order parameter $\phi$ as follows [7, 15-17]: $F(\phi) = \int \left[ \vartheta(\phi) + 0.5\kappa \nabla^2 \phi \right] d\Omega$, where $\vartheta(\phi)$ is the bulk free-energy density, $\kappa \nabla^2 \phi$ accounts for the surface energy, $\Omega$ is the space occupied by the system. For binary fluids, the bulk free-energy density is often chosen to have a double-well form:

$$\vartheta(\phi) = \beta \phi^2 (\phi - 1)^2 \quad \text{or} \quad \vartheta(\phi) = \beta \left( \phi^2 - \phi^{*2} \right)^2, \tag{1}$$

where $\phi^*$ is a constant that corresponding to the equilibrium state, $\phi = \pm \phi^*$. The parameters $\beta$ and $\kappa$ are used to control the surface tension $\sigma$ and the interface thickness $W$. The chemical potential is defined as $\mu_\phi = \vartheta'(\phi) - \kappa \nabla^2 \phi$. Cahn and Hilliard [14] have generalized the time-dependent governing equation of $\phi$. The convective Cahn-Hilliard equation is given as follows [16-21]

$$\partial_t \phi + \boldsymbol{u} \cdot \nabla \phi = \nabla \cdot \left( \theta_M \nabla \mu_\phi \right), \tag{2}$$

where $\theta_M$ is the mobility coefficient. In the phase-field method, the hydrodynamic equations for incompressible multiphase flows are given by [16-21]:

$$\nabla \cdot \boldsymbol{u} = 0, \quad \rho \left( \partial_t \boldsymbol{u} + \boldsymbol{u} \cdot \nabla \boldsymbol{u} \right) = -\nabla p + \nabla \cdot \boldsymbol{\Pi} + \boldsymbol{F}_s + \boldsymbol{G}, \tag{3}$$

where $p$ is the hydrodynamic pressure; $\boldsymbol{\Pi} = \mu \left[ \nabla \boldsymbol{u} + (\nabla \boldsymbol{u})^{\mathrm{T}} \right]$ ($\mu$ is the viscosity) is the viscous stress tensor; $\boldsymbol{G}$ is the body force; and $\boldsymbol{F}_s$ represents the force associated with surface tension. In the literature, the coupled Cahn–Hilliard/Navier–Stokes system (Eqs. (2) and (3)) is referred to as "Model H" [17, 18, 20]. The density $\rho$ in Eq. (3) is taken as a function of the order parameter:

$$\rho = \rho_G + \phi (\rho_L - \rho_G) \quad \text{or} \quad \rho = \rho_G + \frac{\phi + \phi^*}{2\phi^*} (\rho_L - \rho_G), \tag{4}$$

where $\rho_L$ and $\rho_G$ are the densities of liquid and vapor phases at saturation. As pointed out in Ref. [21], since the density $\rho$ is an affine function of the order parameter, the usual mass conservation $\partial_t \rho + \nabla \cdot (\rho \boldsymbol{u}) = 0$ is no longer a consequence of the incompressibility condition ($\nabla \cdot \boldsymbol{u} = 0$) because

$$\partial_t \rho + \nabla \cdot (\rho \boldsymbol{u}) = \frac{d\rho}{d\phi}(\partial_t \phi + \boldsymbol{u} \cdot \nabla \phi) = \frac{d\rho}{d\phi} \nabla \cdot (\theta_M \nabla \mu_\phi). \tag{5}$$

However, by integrating Eq. (5) over the occupied space $\Omega$ and using the divergence theorem, it can be found that the mass of the whole system is conserved when there is no volume diffusive flux across the boundaries ($\theta_M \nabla \mu_\phi$ denotes volume diffusive flux and $\mu_\phi$ is zero in every single-phase region). Actually, within the framework of the phase-field theory with a Cahn-Hilliard-like interface capturing equation, the incompressibility condition and the usual mass conversation cannot be simultaneously satisfied in the interface (the mixing layer) [16-21].

Now we discuss the difference between the recovered momentum equation of the models devised in Refs. [9-11] and the target momentum equation in Eq. (3). As previously mentioned, in these models the velocity field is solved by a pressure distribution function, which is defined as $g_\alpha = c_s^2 f_\alpha + \psi(\rho) \omega_\alpha(0)$, where $f_\alpha$ is the density distribution function, $c_s = \sqrt{RT_0}$, $\psi(\rho) = (p - \rho c_s^2)$, and $\omega_\alpha(\boldsymbol{u}) = w_\alpha \left(1 + u_a + 0.5 u_a^2 - 0.5 u^2/c_s^2\right)$, in which $u_a = (\boldsymbol{e}_\alpha \cdot \boldsymbol{u})/c_s^2$ ($\boldsymbol{e}_\alpha$ are discrete velocity set) and the weights $w_\alpha$ are given by $w_0 = 4/9$, $w_{1-4} = 1/9$, and $w_{5-8} = 1/36$. The evolution equation of $f_\alpha$ is given by [10]

$$\partial_t f_\alpha + \boldsymbol{e}_\alpha \cdot \nabla f_\alpha = -\frac{1}{\tau}\left(f_\alpha - f_\alpha^{eq}\right) + \frac{(\boldsymbol{e}_\alpha - \boldsymbol{u}) \cdot \boldsymbol{F}}{\rho c_s^2} f_\alpha^{eq}, \tag{6}$$

where $\tau$ is the relaxation time, $\boldsymbol{F} = -\nabla \psi + \boldsymbol{F}_s + \boldsymbol{G}$, and $f_\alpha^{eq} = \rho \omega_\alpha(\boldsymbol{u})$. According to Eq. (6), the evolution equation of $g_\alpha$ is given as follows

$$\partial_t g_\alpha + \boldsymbol{e}_\alpha \cdot \nabla g_\alpha = -\frac{1}{\tau}\left(g_\alpha - g_\alpha^{eq}\right) + (\boldsymbol{e}_\alpha - \boldsymbol{u}) \cdot \boldsymbol{F} \omega_\alpha(\boldsymbol{u}) + \omega_\alpha(0)(\partial_t \psi + \boldsymbol{e}_\alpha \cdot \nabla \psi), \tag{7}$$

where $g_\alpha^{eq} = c_s^2 f_\alpha^{eq} + \psi(\rho) \omega_\alpha(0)$. By assuming $\partial_t \psi + \boldsymbol{u} \cdot \nabla \psi = 0$ [22], Eq. (7) can be rewritten as

$$\partial_t g_\alpha + \boldsymbol{e}_\alpha \cdot \nabla g_\alpha = -\frac{1}{\tau}\left(g_\alpha - g_\alpha^{eq}\right) + \left(\boldsymbol{e}_\alpha - \boldsymbol{u}\right)\cdot \boldsymbol{K}, \tag{8}$$

where $\boldsymbol{K} = \left(\boldsymbol{F}_s + \boldsymbol{G}\right)\omega_\alpha(\boldsymbol{u}) - \left(\omega_\alpha(\boldsymbol{u}) - \omega_\alpha(0)\right)\nabla\psi$. Equation (8) is just the evolution equation of $g_\alpha$ adopted in Refs. [9-11].

Through the Chapman-Enskog analysis, the following hydrodynamic equations can be readily derived from Eq. (8) in the low Mach number limit:

$$\frac{1}{\rho c_s^2}\left(\partial_t p + \boldsymbol{u}\cdot\nabla p\right) + \nabla\cdot\boldsymbol{u} = 0, \tag{9}$$

$$\partial_t(\rho\boldsymbol{u}) + \nabla\cdot(\rho\boldsymbol{u}\boldsymbol{u}) = -\nabla p + \nabla\cdot\boldsymbol{\Pi} + \boldsymbol{F}_s + \boldsymbol{G}. \tag{10}$$

For incompressible flows, $\partial_t p$ is very small and $\boldsymbol{u}\cdot\nabla p$ is the order of $O(\mathrm{Ma}^3)$ (Ma is the Mach number). Then the divergence-free condition can be approximately satisfied. However, the left-hand side of Eq. (10) was deemed to be $\rho(\partial_t \boldsymbol{u} + \boldsymbol{u}\cdot\nabla\boldsymbol{u})$ in Refs. [10, 11]. This is the reason why He *et al.* and Lee and Lin claimed that the recovered momentum equation of their models is given by Eq. (3). It is clear that such a transformation is only valid under the condition of $\partial_t\rho + \nabla\cdot(\rho\boldsymbol{u}) = 0$. However, in these models $\partial_t\rho + \nabla\cdot(\rho\boldsymbol{u})$ is nonzero, e.g., in Fakhari and Rahimian's model, $\partial_t\rho + \nabla\cdot(\rho\boldsymbol{u})$ is approximately given by Eq. (5), while in Lee and Lin's model, it takes the following form [11]:

$$\partial_t\rho + \nabla\cdot(\rho\boldsymbol{u}) = \nabla\cdot\left[\lambda(\nabla P - \nabla p)\right], \tag{11}$$

where $\lambda = \mu/(\rho c_s^2)$ and $P$ is the thermodynamic pressure. Since $\partial_t\rho + \nabla\cdot(\rho\boldsymbol{u})$ is nonzero, the recovered momentum equation (10) should be rewritten as follows:

$$\rho(\partial_t\boldsymbol{u} + \boldsymbol{u}\cdot\nabla\boldsymbol{u}) = -\nabla p + \nabla\cdot\boldsymbol{\Pi} + \boldsymbol{F}_s + \boldsymbol{G} + \boldsymbol{F}_a, \tag{12}$$

where $\boldsymbol{F}_a = -\boldsymbol{u}\left[\partial_t\rho + \nabla\cdot(\rho\boldsymbol{u})\right]$. Obviously, an additional term/force $\boldsymbol{F}_a$ is included in the above momentum equation as compared with the target momentum equation. According to the form of $\boldsymbol{F}_a$, several statements can be made. First, it is obvious that, for LB models with different forms of $\partial_t\rho + \nabla\cdot(\rho\boldsymbol{u})$, the additional force will be different. Second, since $\partial_t\rho + \nabla\cdot(\rho\boldsymbol{u})$ is zero in every single-phase region but is nonzero in the interface (the mixing layer), $F_a$ can be viewed as an *interfacial* force. Furthermore, it is believed that, with the increase of the velocity $u$, $F_a$ will gradually play an important role. The influences of $F_a$ can be examined by comparing the results of an original LB model and its corrected model. In this work, we take Fakhari and Rahimian's LB model [9] as the original model considering the Cahn-Hilliard equation is solved in this model. In its corrected version, the additional force is removed by adding $-F_a$ in the forcing term.

Two numerical tests are considered. The first test is droplet splashing on a thin liquid film. This phenomenon can be found in raindrop splashing on a wet surface. In simulations, the second-order accuracy discretization scheme of Eq. (8) is adopted [9, 10] and the widely used isotropic central schemes [11, 23] are applied to calculate the first-order derivatives and the Laplacian. A grid size of $N_x \times N_y = 300 \times 150$ is employed. The height of the liquid film is one-tenths of the entire domain height. The Weber number ($\text{We} = \rho_L U^2 D / \sigma$) is set to be 960, where $\sigma = 10^{-4}$ is the surface tension, $U = 0.04$ is the impact velocity, and $D = 60$ is diameter of the droplet. Two Reynolds numbers ($\text{Re} = UD/\nu_L$) are considered, i.e., $\text{Re} = 48$ and $480$ with $\nu_L = 0.05$ and $0.005$, respectively. The interface thickness $W$, $\rho_L$, and $\rho_G$ are chosen as: $W = 3$, $\rho_L = 1$, and $\rho_G = 0.1$. Figure 1 shows the density contours ($\rho = 0.55$) at $\text{Re} = 48$. The non-dimensional time is defined as $t^* = tU/D$. It is seen that the results of the original and corrected models are nearly the same, which is expected as the Reynolds number is small. Figure 2 presents the results at $\text{Re} = 480$. From Fig. 2(b) we can see that, at $t^* = 1.2$, a pair of secondary droplets has formed from the cusps of the rim, while there are no secondary droplets in Fig. 2(a). As time goes on, two pairs of secondary droplets can be found in Fig. 2(b). At the same time, a pair of secondary droplets can also be observed in Fig. 2(a). However, the size of the droplets is smaller and their position is lower than that of the first pair of secondary droplets in Fig. 2(b). In fact, this is because that the droplets are surrounded by downward additional forces, which can be seen clearly from Fig. 3, where the corresponding *y*-component of $\mathbf{F}_a$ is shown. That is to say, the secondary droplets in Fig. 2(a) are dragged down by the additional interfacial force. To sum up, the density contours in Fig. 2(b) are smoother and reasonably confirm the mechanism of droplet splashing summarized in Ref. [24].

The second test is the evolution of a falling droplet under gravity. In simulations, a grid size of $N_x \times N_y = 150 \times 450$ is adopted. The periodic boundary condition is applied in the *x*-direction. The Eötvös number defined by $\text{Eo} = g(\rho_L - \rho_G)D^2/\sigma$ ($g$ is the gravitational acceleration) is set to be $\text{Eo} = 17$ with $\sigma = 0.002$, $D = 32$, $\rho_L = 5$, and $\rho_G = 1$. The interface thickness $W$ and the viscosity $\nu_L$ are chosen as $W = 3$ and $\nu_L = 0.025$. Figure 4 shows the evolution of the droplet as it falls. From the figure we can see that, up to $t = 10^4$, there are nearly no differences between the results of the original and corrected models. However, from $t = 1.2 \times 10^4$, it is observed that the droplet in Fig. 4(a) gradually gets less stretched in the radial direction and its position is lower than the corresponding position in Fig. 4(b). This means that, with the increase of the velocity, the additional interfacial force surrounding the droplet (not shown here) begins to accelerate the falling process of the droplet. Meanwhile, the stretch of the droplet is also limited by the additional force. To see this point more clearly, a comparison of spread radius between the original and corrected models is made in Fig. 5. Evident differences can be observed from $t = 10^4$. In particular, at $t = 1.9 \times 10^4$, the difference is 4.1 l.u. (lattice units), which is about 26 percents of the initial radius.

In summary, we have shown that in most of previous incompressible multiphase LB models an additional interfacial force is included in the recovered momentum equation. The effects of the additional force have been numerically investigated by comparing the results of an original model and its corrected model that does not have the additional force. In the problem of droplet splashing on a thin liquid file, it is found that, with the increase of the Reynolds number, the additional force has an important influence on the interface. Particularly, the formation and evolution of secondary droplets are greatly affected. In the evolution of a falling droplet under gravity, the additional force is found to increase the falling velocity and limit the stretch of the droplet.

This work was supported by the Engineering and Physical Sciences Research Council of United Kingdom under Grant Nos. EP/I000801/1 and EP/I012605/1.

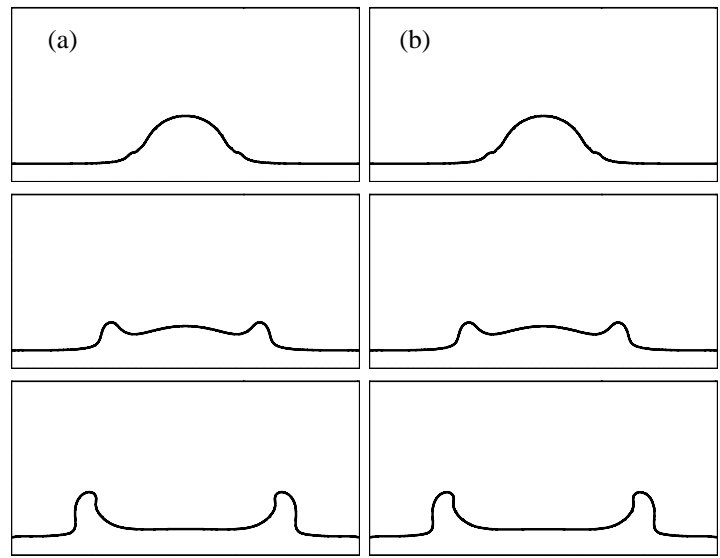

FIG. 1. Droplet splashing at $Re = 48$: (a) original and (b) corrected. From top to bottom: $t^* = 0.4$, $1.0$, and $2.0$.

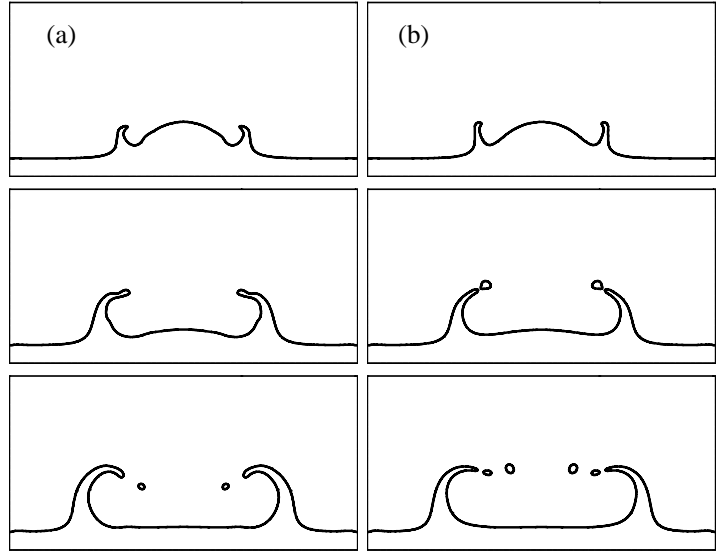

FIG. 2. Droplet splashing at $Re = 480$: (a) original and (b) corrected. From top to bottom: $t^* = 0.6$, $1.2$, and $1.8$.

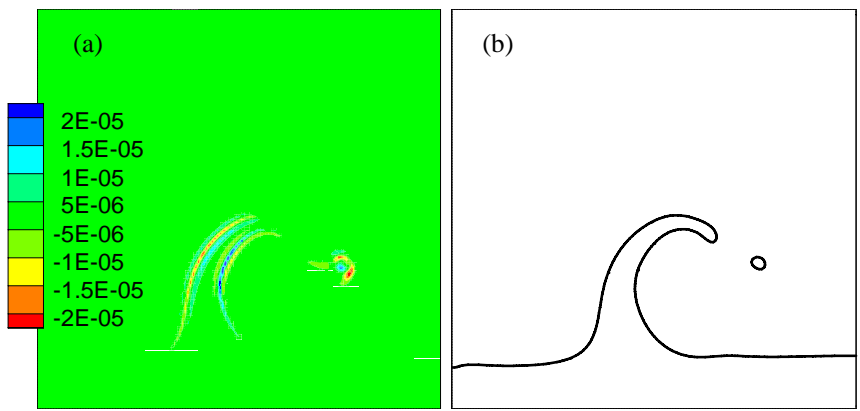

FIG. 3. (a) The *y*-component of $F_a$ at $t^* = 1.8$; (b) the corresponding density contour.

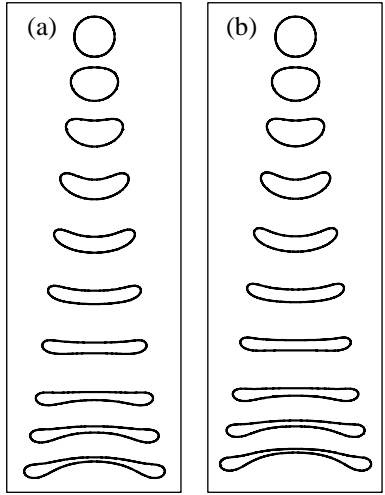

FIG. 4. Evolution of falling droplet under gravity ($t/10^3 = 0.1, 4, 6, \cdots, 16, 17.5, 19$):
(a) original and (b) corrected.

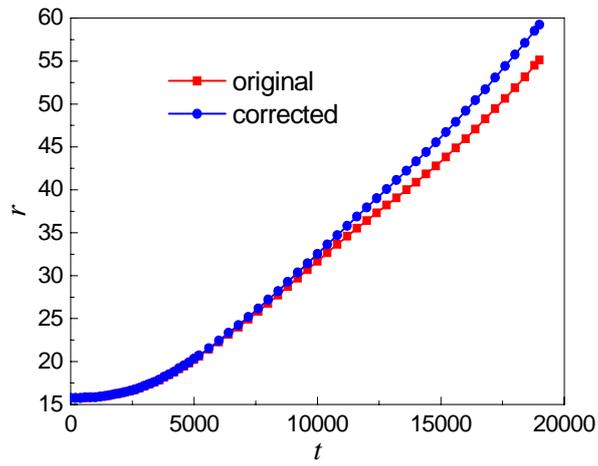

FIG. 5. Comparison of spread radius between the original and corrected models.